\def\preprint{1}
\ifnum\preprint=9
    \documentclass[11pt,draft,a4paper]{article} 
\fi 
\def\masterthesis{0}
\def\bigfont{0} 
\def\cryptology{0} 
\def\sigconf{0} 
\def\llncs{0} 
\def\lipics{0} 
\def\quantumjournal{0}

\def\draft{0}
\def\anonymous{0}
\def\shownomenclature{1} 
\def\smallbib{1} 
\def\toc{0} 
\ifnum
    \ifnum\preprint=1 1\else\ifnum\cryptology=1 1\else 0\fi\fi
=1 
    \ifnum\draft=1
       \documentclass[11pt,draft,a4paper]{article}
       \pdfoutput=1
    \else
        \documentclass[11pt,final,a4paper]{article}
        \pdfoutput=1
    \fi
    \usepackage{lmodern}
\fi
\ifnum\masterthesis=1
    \documentclass[11pt,final,a4paper,titlepage]{article}
    \usepackage{lmodern}
\fi
\ifnum\sigconf=1
        \documentclass[sigconf,final]{acmart}
\fi
\ifnum\bigfont=1
    \documentclass[final,17pt]{extarticle} \usepackage{fullpage}
\fi
\ifnum\llncs=1
    \documentclass[runningheads]{llncs}
\fi
\ifnum\lipics=1
    \input{lipics_header}
\fi
\ifnum\quantumjournal=1
  \documentclass[a4paper,onecolumn,11pt 
  ]
  {quantumarticle}
  \pdfoutput=1

\fi

\ifnum\toc=1\ifnum\lipics=0
    \usepackage[nottoc]{tocbibind} 
\fi\fi
\ifnum\lipics=0\ifnum\sigconf=0\usepackage{authblk}\fi\fi 
\ifnum\sigconf=0
    \usepackage{silence} 
    \usepackage{amssymb} 
\fi
\ifnum\lipics=0 
    \usepackage[modulo]{lineno}
\fi

\ifnum\llncs=1

\fi \usepackage{amsthm}

\usepackage{amsmath,amsfonts,latexsym,color,paralist,url,ifdraft,mathrsfs,comment,bbold,booktabs,ifthen,wrapfig}

\usepackage[utf8]{inputenc} \usepackage[autostyle=false, style=english]{csquotes} \MakeOuterQuote{"}

\usepackage[T1]{fontenc}
\ifnum\sigconf=0
    \ifnum \shownomenclature=1
        \usepackage[refpage,prefix]{nomencl}
         
        \makenomenclature
        
  \makeatletter
        \def\thenomenclature{%
          \providecommand*{\listofnlsname}{\nomname}%

              \section{\nomname}
              \label{sec:nomenclature}    
              \if@intoc\addcontentsline{toc}{section}{\nomname}\fi%

          \nompreamble
          \if@nomentbl
            \let\itemOrig=\item
            \def\item{\gdef\item{\\}}%
            \expandafter\longtable\expandafter{\@nomtableformat}
          \else
            \list{}{%
              \labelwidth\nom@tempdim
              \leftmargin\labelwidth
              \advance\leftmargin\labelsep
              \itemsep\nomitemsep
              \let\makelabel\nomlabel}%
          \fi
        }
        \makeatother 
    \fi
    \ifnum\lipics=0    
        \usepackage[draft=false,pageanchor,hypertexnames=false]{hyperref}
        \usepackage[final]{graphicx}
    \fi
    \usepackage{doi}
\fi

\ifdraft{\usepackage{showlabels}}{} 

\ifdraft{\newcommand{\authnote}[3]{{\color{#3} {\bf  #1:} #2}}}{\newcommand{\authnote}[3]{}}

\ifdraft{\linenumbers}{}

\usepackage{algorithm,algorithmicx,algpseudocode}
\usepackage[normalem]{ulem}

\usepackage{tikz}
\usetikzlibrary{arrows,chains,matrix,positioning,scopes}
\makeatletter
\tikzset{join/.code=\tikzset{after node path={%
\ifx\tikzchainprevious\pgfutil@empty\else(\tikzchainprevious)%
edge[every join]#1(\tikzchaincurrent)\fi}}}
\makeatother
\tikzset{>=stealth',every on chain/.append style={join},
         every join/.style={->}}
\tikzstyle{labeled}=[execute at begin node=$\scriptstyle,
   execute at end node=$]
\usepackage[capitalise]{cleveref} 

\usepackage{autonum}
\ifthenelse{
\(\equal{\cryptology}{1}\OR
\(\equal{\preprint}{1} \OR \equal{\masterthesis}{1}\) \OR \(\equal{\bigfont}{1} \OR \equal{\quantumjournal}{1} \)
\) 
}
{
    \newtheorem{theorem}{Theorem} 
    \newtheorem{lemma}{Lemma}
    \newtheorem{corollary}{Corollary}
    \newtheorem{proposition}{Proposition}
    
    \newtheorem{example}{Example}
    
    \newtheorem{claim}{Claim}
    \newtheorem{fact}{Fact}
    \newtheorem*{theorem*}{Theorem}
    \newtheorem*{lemma*}{Lemma}
    \newtheorem*{corollary*}{Corollary}
    \newtheorem*{proposition*}{Proposition}
    \newtheorem*{claim*}{Claim}
    \theoremstyle{definition}
    \newtheorem{openproblem}{Open Problem}
    \theoremstyle{remark}
    \newtheorem{remark}[theorem]{Remark}

    \theoremstyle{plain}
    
}{}  
\ifnum\sigconf=1

  \newtheorem*{theorem*}{Theorem}
  \newtheorem*{lemma*}{Lemma}
  \newtheorem*{corollary*}{Corollary}
  \newtheorem*{proposition*}{Proposition}
  \newtheorem*{claim*}{Claim}
  \theoremstyle{definition}

\fi

\theoremstyle{plain}
\usepackage{float}
\newfloat{game}{htbp}{gam}
\floatname{game}{Game}
\crefname{game}{Game}{Games}

\newcommand{\ab}[1]{}  

\ifnum\masterthesis=1
    \usepackage{setspace}
    \usepackage{fancyhdr}
\fi

\newcommand{\onote}[1]{\authnote{Or}{#1}{blue}}

\usepackage{pgfplots}
\pgfplotsset{compat=1.18}
\usepackage{xspace}
\DeclareMathAlphabet{\mathpzc}{OT1}{pzc}{m}{it}

\newtheorem{assumption}{Assumption}
\crefname{assumption}{Assumption}{Assumptions}
\Crefname{assumption}{Assumption}{Assumptions}
\begin{document}
\ifnum\masterthesis=0
    \title{Useful Work Can Increase Energy Use:\\
          Demand Elasticity and Proof of Useful Work}
\fi

\ifnum\anonymous=0
    \ifnum\preprint=1
        \author[1]{Isaiah Hull}
        \affil[1]{FirstQFM}
        \author[2]{Or Sattath}
        \affil[2]{Computer Science Department, Ben-Gurion University of the Negev}
    \fi
    \ifnum\cryptology=1
        \author{XXX}
        \affil{Department of Computer Science, Ben Gurion University of the Negev, Beersheba, Israel\\
        XXX@post.bgu.ac.il}
        \author{Or Sattath}
        \affil{Department of Computer Science, Ben Gurion University of the Negev, Beersheba, Israel\\
        sattath@bgu.ac.il}
    \fi
    \ifnum\sigconf=1
        \author{Or Sattath}
            \affiliation{%
              \institution{Department of Computer Science, Ben Gurion University of the Negev}
              \city{Beersheba}
              \country{Israel}}
        \email{sattath@bgu.ac.il}

        \author{XXX}
            \affiliation{%
              \institution{Department of Computer Science, Ben Gurion University of the Negev}
              \city{Beersheba}
              \country{Israel}}
        \email{XXX@bgu.ac.il}
    \fi
    \ifnum\llncs=1
        %
        \author{Or Sattath\inst{1}\orcidID{0000-0001-7567-3822} \and
        Second Author\inst{2,3}\orcidID{1111-2222-3333-4444} \and
        Third Author\inst{3}\orcidID{2222--3333-4444-5555}}
        \authorrunning{O. Sattath}
        %
        \institute{Department of Computer Science, Ben Gurion University of the Negev, Beersheba, Israel 
        \email{sattath@bgu.ac.il}\\
        \url{http://www.springer.com/gp/computer-science/lncs} \and
        ABC Institute, Rupert-Karls-University Heidelberg, Heidelberg, Germany\\
        \email{\{abc,lncs\}@uni-heidelberg.de}}
    \fi
    \ifnum\quantumjournal=1
        \author[1]{Andrea Coladangelo}
        \affil[1]{Computing and Mathematical Sciences,	
        Caltech}
        \author[2]{Or Sattath}
        \orcid{0000-0001-7567-3822}
        \email{sattath@bgu.ac.il}

        \affil[2]{Computer Science Department, Ben-Gurion University of the Negev}
        
    \fi
\else
    \ifnum\lipics=0
        \author{}
    \fi
\fi

\ifthenelse{\equal{\masterthesis}{0} \AND \equal{\sigconf}{0}}{
    \maketitle
} 
{}        
\ifnum\masterthesis=1
    \include{masterthesis_header}
\else 
    \ifnum\sigconf=0
        \begin{abstract}
    \fi
\fi
\ifnum\sigconf=0
    \ifnum\masterthesis=0
            A serious criticism of Bitcoin's Proof-of-Work is its enormous electricity consumption. 
            Proof-of-Useful-Work (PoUW), introduced by Ball, Rosen, Sabin, and Vasudevan (2017), in which mining performs useful computations, was suggested as an alternative to reduce electricity expenditure, with matrix multiplication (MatMul) as a recent concrete instantiation by Komargodski and Weinstein (2026).

            Perhaps surprisingly, we show that switching to PoUW may \emph{increase} the total electricity consumption, echoing Jevons' paradox.
            Intuitively, the PoUW mechanism acts as a subsidy that lowers the effective price of the useful work.
            Recall that demand is elastic if a discount of $x$ percent in the price leads to an increase of more than $x$ percent in the quantity demanded. 
            If demand is elastic, the resulting increase in quantity demanded leads to a net increase in total electricity consumption; if inelastic, total electricity consumption decreases. Even then, the electricity consumption due to PoUW never vanishes, since the mining subsidy is spent on useful work that would not have been produced without the subsidy.
            This result is proved in a simple model with electricity-only costs, no PoUW overhead, and an unchanged mining subsidy. The analysis extends to less restrictive settings with adjusted thresholds.
    \fi
\fi 
\ifnum\masterthesis=0
    \ifnum\sigconf=0
        \end{abstract}
    \fi
\fi

\ifnum\sigconf=1
    \begin{abstract}
        Abstract of SigConf goes here.
    \end{abstract}
    \maketitle
\fi

\ifnum\toc=1
    \ifnum\llncs=1
        \setcounter{secnumdepth}{3}
        \setcounter{tocdepth}{3}
    \fi
    \tableofcontents 
\fi

\onote{We agreed to submit it to 1) Economics Letters. 2) Finance Research Letters.}
\section{Introduction} 
\label{sec:introduction} 
A major drawback of Bitcoin's Proof-of-Work (PoW) is ``its energy-intensive nature and exorbitant environmental costs (consuming over 2\% of annual electricity in the US alone)''~\cite{KW26}.
Indeed, reducing energy consumption was one of the primary stated motivations for Ethereum's 2022 transition to Proof-of-Stake (PoS): the Ethereum Foundation switched ``because it is \ldots{} less energy-intensive''~\cite{Eth22}. This transition reduced Ethereum's electricity consumption by over 99.988\%~\cite{CCRI22}.

 Komargodski and Weinstein~\cite{KW26} argue that Proof-of-Useful-Work (PoUW), a notion introduced by Ball et al.~\cite{BRS+17}, ``nearly eliminates the energy waste
of Bitcoin mining.'' Indeed, one would na\"ively expect that the transition from PoW to PoUW would reduce the total electricity consumption. 

\subsection{Our Contribution}
Our main result shows that the na\"ive intuition mentioned above is incorrect: under certain conditions, the total electricity consumption will \emph{increase} when transitioning to PoUW. Recall that demand is elastic if a discount of $x$ percent in the price increases the quantity demanded by more than $x$ percent (see \cref{eq:elasticity} for the formal definition).
Informally, we show that electricity consumption increases if demand for the useful resource is elastic, and decreases if it is inelastic. In either case, it is not eliminated, in stark contrast to the near-total reduction achieved by the PoS transition, and to the near elimination suggested by Komargodski and Weinstein~\cite{KW26}.

Our result can be understood through the lens of the \emph{rebound effect} and, in particular, \emph{Jevons' paradox}~\cite{Jev66,Sor09}. In 1866, Jevons observed that as steam engines became more fuel-efficient, Britain's total coal consumption rose rather than fell. Two effects compete: the efficiency gain directly reduces coal use per unit of output, but lower operating costs also stimulate demand for the goods those engines produce---factory output, textiles, transport, and so on---raising total production and hence total coal use. If demand is inelastic, the efficiency gain dominates and coal consumption falls, though by less than the na\"ive expectation (\emph{partial rebound}). If demand is elastic, the demand expansion outweighs the efficiency gain and total coal consumption rises---the \emph{backfire} case that Jevons originally described.

Our result is analogous, though the mechanism differs. In the classical rebound framework, a technological improvement provides a fixed, exogenous per-unit reduction in resource use, uniformly lowering costs for every unit of output. Here, miners abandon PoW, saving the total mining reward $\$R$ (per unit time) in electricity, and this $\$R$ is redirected as a total subsidy to those performing useful work. Because the subsidy is shared across all units produced, the per-unit benefit is endogenous: it shrinks as production expands, rather than remaining a fixed efficiency gain. Nevertheless, the qualitative outcome mirrors the rebound framework: when demand is inelastic, electricity consumption falls but the demand rebound prevents full elimination (partial rebound); when demand is elastic and the mining reward $R$ is small relative to the size of the useful-work market, the increase in useful-work production consumes more electricity in aggregate than the $\$R$ saved by stopping PoW---that is, there is \emph{backfire}. These conclusions are made precise in \cref{prop:linear_demand_condition_for_increased_electricity,prop:general_condition_for_increased_electricity}. Both results rely on the modeling assumptions specified in \cref{sec:model}---in particular, that all costs are spent on electricity---with this assumption relaxed in \cref{sec:non_energy_costs}.

\subsection{Related Work}
A substantial body of work has estimated and analyzed Bitcoin's electricity consumption and its environmental footprint~\cite{Vri18,SKG19,VGKS22}, as well as that of other cryptocurrencies~\cite{GKS20}; ongoing estimates are maintained by the \href{https://ccaf.io/cbnsi/cbeci}{Cambridge Bitcoin Electricity Consumption Index (CBECI)}.

Bar-On, Komargodski, and Weinstein~\cite{BKW25} study PoUW from the miners' side. Fiat et al.~\cite{FKKP19} characterize the Nash equilibrium of PoW mining when miners face heterogeneous costs, such as differing electricity prices. Bar-On et al.\ extend this model to the PoUW setting, in which miners additionally earn external rewards for performing useful computation, such as AI training and inference. They characterize the resulting equilibrium among a finite set of miners, show that miners with access to external rewards may find it profitable to concentrate their useful tasks in a single block, and study the consequences for decentralization, measured as the Shannon entropy of the distribution of computational effort among miners. Their analysis takes the external reward per unit of useful work as given. In our model, by contrast, competition passes the mining reward through to the consumers of the useful resource as a lower price, and it is the resulting demand response that determines the aggregate electricity consumption. Bar-On et al.\ also argue that their ``results support the potential of PoUW to retain PoW-level security with lower net costs and environmental effects, thanks to computational reuse.''
Our results challenge the claimed environmental benefits: when demand for the useful resource is elastic, the transition to PoUW \emph{increases} total electricity consumption. Even when demand is inelastic, the reduction is far from dramatic, since the mining reward is still spent on electricity, through useful work that is produced only because of the subsidy.

The work closest to ours is that of Pass~\cite{Pas26}, who also studies the market equilibrium of a PoUW blockchain. Pass examines it through the lens of security, whereas we focus on electricity consumption. The two works were carried out concurrently and independently, with one exception: \cref{sec:overhead}, which we wrote after his work appeared and in reaction to it, as discussed below. Pass develops a competitive-equilibrium model in which compute is allocated among pure mining, pure useful work (ML inference), and \emph{duplex} work that produces both, with multiplicative overheads relative to each pure activity. Equilibria are parameterized by the \emph{token-inference ratio} $\theta$, the dollar value of block rewards divided by inference spending: for each $\theta$ there is a unique equilibrium. Pass argues that $\theta$ is a monetary constant independent of the technology. Depending on the overheads and $\theta$, the economy is in one of three regimes: \emph{Bitconia}, which reduces to classical PoW; \emph{Fortessia}, in which duplex work replaces pure mining, raising security while leaving the price and quantity of inference unchanged; and \emph{Duplexia}, in which block rewards subsidize inference, pushing its price below cost and expanding its supply. His bottom line concerns security. PoUW is often met with what Pass calls a seemingly decisive straw-man objection: if the work is useful, an attacker can be ``paid to attack'' by selling the useful output, so the cost of attacking collapses. Contrary to this argument, he shows that once equilibrium prices are taken into account, the economic cost of a majority attack equals half the block-reward value in every regime, exactly as in PoW.

The two models are closely related. With no duplex overhead and a fixed dollar reward rather than a fixed $\theta$, his zero-profit condition reduces to our break-even condition in \cref{eq:supply_curve}, and the economy is always in Duplexia. The two works, however, shine their light on different sides of the same equilibrium. Pass emphasizes the positive ones. Beyond the cost of attack, which does not decrease, he shows that in Duplexia the block rewards act as rebates on inference prices, generating additional socially useful computation that would not arise without the blockchain, and that this expansion persists even when demand is inelastic. He then quantifies the net change in social welfare from the transition, accounting for the deadweight loss and the duplex overhead on the additional computation. This comparison is favorable to PoUW only up to a point: under isoelastic demand with elasticity $\varepsilon>1$, Duplexia yields strictly higher social welfare than classical PoW whenever $\theta<1/(\varepsilon-1)$, whereas for larger $\theta$ the deadweight loss and overhead on the expanded computation outweigh the savings, and social welfare falls below that of PoW. Our work points to a negative side of the very same expansion: the additional useful computation consumes electricity, and when demand is elastic, it consumes more than the hashing it replaces.
The two effects, the demand elasticity studied here and the PoUW overhead studied by Pass, interact, and a reader may wish to understand them together. For such readers, \cref{sec:overhead} extends our analysis to take the PoUW overhead into account, using the same conventions as Pass.

The closest analogue outside cryptoeconomics is the energy-economics literature on the rebound effect and \emph{backfire}. Greening, Greene, and Difiglio survey direct, indirect, and economy-wide rebound channels created when efficiency improvements lower the effective price of energy services~\cite{Gre00}. Saunders gives a canonical theoretical formulation of the Khazzoom--Brookes postulate, identifying conditions under which efficiency improvements can raise aggregate energy use~\cite{Sau92}; Sorrell reviews the evidence and debate surrounding such backfire claims~\cite{Sor09}. Borenstein provides a microeconomic framework that decomposes rebound into substitution and income effects~\cite{Bor15}, which is especially close in spirit to our price-elasticity analysis. Closer to the computing setting, Galvin argues that information and communication technology (ICT) and electronics can exhibit large rebound effects because efficiency gains lower the cost of computation and induce new applications and structural changes in demand~\cite{Gal15}. Our setting differs from this literature in that the price reduction is generated not by an exogenous efficiency improvement, but by reallocating a fixed mining reward as an endogenous subsidy whose per-unit magnitude falls as output expands.

We are not aware of an empirical estimate specifically for the demand elasticity of matrix multiplication (MatMul). The closest historical cloud-market proxy we found is the study of Jiang, Lee, and Zomaya, who estimate a 2013 elasticity of 1.2 for early AWS EC2 public-cloud demand~\cite[p.~29]{JLZ16}.
More recent cloud-demand studies point in the opposite direction: Ref.~\cite[Table~6]{JPW26} estimates average own-price elasticities between 0.629 and 0.942 across virtual-machine groups, with only one large-customer category slightly above unit elasticity, while Brand et al.~\cite{BCLM25} estimate an elasticity of about 0.5 in a spot/on-demand setting. These papers are still cloud virtual-machine studies rather than estimates for compute in general, and none directly identify MatMul demand elasticity.
 
\section{The Model}
\label{sec:model}
\subsection{Setup}
 
\paragraph{Comparison Framework.}
The model covers any useful-work resource; for concreteness, it is useful to consider MatMul-based PoUW~\cite{KW26}. The goal is to compare a cryptocurrency using PoW with the same cryptocurrency after a transition to PoUW, holding all other protocol parameters fixed, including the cryptocurrency's price\footnote{Pass~\cite[Sections~1.3 and~1.6, Appendix~A]{Pas26} adopts a different comparison: he holds fixed the token--inference ratio $\theta$, defined as the dollar value of block rewards divided by total inference spending. With token issuance fixed in tokens, the token price and hence the dollar value of the mining reward adjust with equilibrium inference spending. He motivates the fixed ratio using a monetary model linking token demand to inference settlement, with fixed velocity, circulation, and hoarding parameters. This choice does not affect the backfire threshold of \cref{prop:linear_demand_condition_for_increased_electricity}: with no overhead and linear demand, total electricity consumption under a fixed $\theta$ is $1+\theta$ times the spending on the useful resource, both before and after the transition, so it increases if and only if that spending does; with $\theta=R$, as at our pre-PoUW equilibrium, this happens if and only if $\gamma>1+R$.} and total mining reward $R$ per unit time. Under PoW, this reward is spent on block production through hashing. For the useful-work market, we use a standard supply-and-demand description: users demand the useful resource, such as MatMul, while producers supply it. Under PoUW, the same total reward $R$ is paid as a subsidy to producers of useful work, reducing the effective unit price faced by users.

\paragraph{Demand.} Let $Q_D(P)$ denote the quantity of the useful resource demanded per unit time at consumer price $P$. The (point) elasticity of demand at price $P$ is defined as (see~\cite[Section~2.4]{PR17} and~\cite{Man18})
\begin{equation}
      \varepsilon(P) = -\frac{P}{Q_D(P)} \cdot \frac{dQ_D(P)}{dP}.
\label{eq:elasticity}
\end{equation} 

 The leading minus sign makes $\varepsilon(P)$ nonnegative for goods obeying the law of demand; demand is \emph{elastic} at $P$ if $\varepsilon(P)>1$ and \emph{inelastic} if $\varepsilon(P)<1$.

\paragraph{Supply and Normalization.} We write supply curves in inverse form, with price as a function of quantity. Before the transition, the useful resource is supplied perfectly elastically at unit cost $\$1$, so the pre-PoUW supply curve is
\[
    P_S^{\mathrm{PoW}}(Q)=1.
\]
We also choose quantity units so that the pre-PoUW equilibrium satisfies $Q_D(1)=1$. With both price and quantity normalized to one, the elasticity at this equilibrium simplifies to $\varepsilon(1)=-Q_D'(1)$.
 
\paragraph{Costs.} All costs incurred in PoW mining, in MatMul production, and in future MatMul-based PoUW mining are assumed to be spent on electricity (hardware and other non-energy costs are negligible); \cref{sec:non_energy_costs} relaxes this assumption.

\subsection{Equilibrium under PoUW}
\label{sec:equilibrium}
After the transition, producers receive the consumer price $P$ together with a per-unit subsidy. If the total amount produced is $Q$, the fixed subsidy pool $R$ is spread across these units, so the subsidy per unit is $R/Q$. A producer breaks even when
\[
    P+\frac{R}{Q}=1.
\]
Competition passes the subsidy through as a lower consumer price, as expressed by the break-even equation above. The post-PoUW supply curve is
\begin{equation}
\label{eq:supply_curve}
    P_S^{\mathrm{PoUW}}(Q)=1-\frac{R}{Q}. 
\end{equation}

The equilibrium $(Q^*, P^*)$ is the point lying on both this curve and the demand curve, $Q^* = Q_D(P^*)$. Writing the discount as $\delta := 1 - P^*$, the supply relation, see \cref{eq:supply_curve}, becomes $\delta = R/Q^*$, and substituting $Q^* = Q_D(1-\delta)$ gives
\begin{equation}
\label{eq:condition_on_delta}
\delta = \frac{R}{Q_D(1-\delta)}.
\end{equation}

The supply curve, see \cref{eq:supply_curve}, reaches a price of zero at $Q=R$, so an equilibrium with a positive price exists if and only if $R<Q_D(0)$. We assume this throughout.

\begin{assumption}
\label{asm:positive_price}
    The mining reward is smaller than the quantity consumers would take for free: $R<Q_D(0)$. For the linear demand of \cref{eq:linear_demand}, this condition is $R<1+\gamma$.
\end{assumption}

\section{Main Results}
\label{sec:main_results}
\subsection{Linear Demand} 

We now consider a general linear demand curve that satisfies $Q_D(1)=1$.
Any such linear demand can be written as
\begin{equation}
\label{eq:linear_demand}
Q_D(P) = 1 + \gamma \cdot (1-P),
\end{equation}
for some parameter $\gamma > 0$. Note that $\gamma = \varepsilon(1)$, the price elasticity of demand at the pre-PoUW equilibrium; the elasticity of demand at other prices is not necessarily $\gamma$: a linear demand curve is \emph{not} isoelastic.

\begin{proposition}
    In the linear demand model with $0<R<1+\gamma$, the total electricity consumption grows when transitioning to PoUW if and only if $\gamma > 1 +R$.
    \label{prop:linear_demand_condition_for_increased_electricity}
\end{proposition}
\begin{proof}
    
Substituting the linear demand into \cref{eq:condition_on_delta}, we obtain
\begin{equation}
\label{eq:linear_delta_condition}
    \delta = \frac{R}{1 + \gamma \delta}.
\end{equation}



The economically meaningful root ($\delta>0$) is
\[
\delta
=
\frac{
    -1 + \sqrt{1 + 4 \gamma R}
}{
    2 \gamma
}.
\]

The resulting total production after introducing PoUW is
\begin{equation}
\label{eq:linear_post_pouw_quantity}
Q_D(1-\delta)
= 1 + \gamma \delta
= 1 + \gamma \cdot
    \frac{
        -1 + \sqrt{1 + 4 \gamma R}
    }{
        2 \gamma
    }
= \frac{1 + \sqrt{1 + 4 \gamma R}}{2}.
\end{equation}

Before the transition, useful-work production consumes $1$ unit of electricity, and PoW mining consumes an additional $R$, giving a total of $1+R$. After the transition, all mining consists of useful-work production. Since each unit of useful work costs one unit of electricity and PoUW introduces no overhead, total electricity consumption equals the equilibrium quantity produced, namely $\frac{1+\sqrt{1+4\gamma R}}{2}$.
Simple algebra shows that the electricity consumption grows if and only if $\gamma>1+R$.

\end{proof}
\begin{remark}
    The interesting regime for PoUW is when $R$ is much smaller than the original expenditure on the useful resource, which is normalized to $\$1$. In this regime, the condition for an increase in electricity consumption is approximately $\gamma>1$, namely, that demand is elastic at the pre-PoUW equilibrium.\end{remark}

\begin{example}[A linear-demand backfire case]
\label{ex:linear_demand}
Consider the linear demand curve $Q_D(P)=1+3(1-P)$ and the mining reward $R=\tfrac{1}{4}$.
Here $\gamma=3$, and since $\gamma > 1 + R = \tfrac{5}{4}$, \cref{prop:linear_demand_condition_for_increased_electricity} predicts that total electricity consumption increases after the transition to PoUW.  
The equilibrium discount is
\[
    \delta = \frac{-1 + \sqrt{1 + 4 \cdot 3 \cdot \tfrac{1}{4}}}{2 \cdot 3}
           = \frac{1}{6}.
\]
Thus the consumer price falls from $1$ to
\[
    P^* = 1-\delta = \tfrac{5}{6},
\]
and demand rises from $1$ to
\[
    Q^* = Q_D\!\left(\tfrac{5}{6}\right)
    = 1 + 3 \cdot \tfrac{1}{6}
    = \tfrac{3}{2}.
\]
Total electricity consumption changes from
\[
    1 + \tfrac{1}{4} = \frac{5}{4}
    \quad\longrightarrow\quad
    \frac{3}{2},
\]
an increase of $\tfrac{1}{4}$, or $20\%$, despite the elimination of PoW hashing.
\Cref{fig:example_demand_supply} illustrates the equilibrium shift.
\end{example}
     
\pgfplotsset{
  example panel eq/.style={
    x=3cm, y=3cm,                 
    xmin=0, xmax=2.5,
    ymin=0, ymax=1.45,
    xlabel={$Q$}, ylabel={$P$},
    axis lines=left,
    xlabel style={at={(axis description cs:1,0)}, anchor=north},
    ylabel style={at={(axis description cs:-0.08,0.5)}, anchor=south}, 
    tick label style={font=\small},
    label style={font=\small},
    title style={font=\small\bfseries, yshift=2pt},
    clip=false,
    xtick={1, 1.5}, xticklabels={$1$, $\tfrac{3}{2}$},
    ytick={0.8333, 1, 1.3333}, yticklabels={$\tfrac{5}{6}$, $1$, $\tfrac{4}{3}$},
  }
}
\begin{figure}[htbp]
\centering
\begin{tikzpicture}
\begin{axis}[
  example panel eq,
  xtick={1}, xticklabels={$1$},
  ytick={1, 1.3333}, yticklabels={$1$, $\tfrac{4}{3}$},
  title={\textbf{(a) Before (PoW)}},
]
  \fill[gray!25] (axis cs:0,0) rectangle (axis cs:1,1);
  \addplot[thick, domain=0:2.3, samples=2] {4/3 - x/3};
  \node[right, font=\small] at (axis cs:2.3, 0.567)
    {$D\colon P=\tfrac{4}{3}-\tfrac{Q}{3}$};
  \addplot[red!70!black, semithick, domain=0:2.3, samples=2] {1};
  \node[right, font=\small] at (axis cs:2.3, 1) {$P_S^{\mathrm{PoW}}(Q)=1$};
  \node[right, font=\small] at (axis cs:2.3, 1) {\phantom{$P_S^{\mathrm{PoUW}}(Q)=1-\tfrac{1}{4Q}$}};
  \draw[dashed, gray] (axis cs:1,0) -- (axis cs:1,1);
  \draw[dashed, gray] (axis cs:0,1) -- (axis cs:1,1);
  \addplot[only marks, mark=*, mark size=2pt,
    mark options={fill=black, draw=black}] coordinates {(1,1)};
  \node[above=2pt, font=\small] at (axis cs:1, 1) {$A$};
  \node[font=\small] at (axis cs:0.5, 0.5) {$1$};
  \node[below=20pt, font=\small, align=center] at (axis cs:1.25, 0)
    {$E_{\mathrm{total}} = E_{\text{useful resource}} + E_{\text{PoW}}= 1 + \tfrac{1}{4} = \tfrac{5}{4}$};
\end{axis}
\end{tikzpicture}

\vspace{1.2em}
  
\begin{tikzpicture}
\begin{axis}[
  example panel eq,
  xtick={0.25, 1, 1.5}, xticklabels={$\tfrac{1}{4}$, $1$, $\tfrac{3}{2}$},
  title={\textbf{(b) After (PoUW)}},
]
  \fill[gray!25] (axis cs:0,0) rectangle (axis cs:1.5, 1);
  \node[font=\small] at (axis cs:0.75, 0.3) {$\tfrac{3}{2}$};
  \addplot[thick, domain=0:2.3, samples=2] {4/3 - x/3};
  \node[right, font=\small] at (axis cs:2.3, 0.567)
    {$D\colon P=\tfrac{4}{3}-\tfrac{Q}{3}$};
  \addplot[red!70!black, thick, domain=0.25:2.3, samples=80] {1 - 1/(4*x)};
  \node[right, font=\small] at (axis cs:2.3, 0.891)
    {$P_S^{\mathrm{PoUW}}(Q)=1-\tfrac{1}{4Q}$};
  \addplot[gray!60, dashed, semithick, domain=0:2.3, samples=2] {1};
  \addplot[only marks, mark=o, mark size=2pt,
    mark options={draw=gray!50, fill=white}] coordinates {(1,1)};
  \node[above right=-1pt, font=\small] at (axis cs:1, 1) {$A$};
  \addplot[only marks, mark=*, mark size=2pt,
    mark options={fill=black, draw=black}] coordinates {(1.5, 0.8333)};
  \node[below left=-1pt, font=\small] at (axis cs:1.5, 0.8333) {$B$};
  \draw[->, semithick] (axis cs:1.07, 0.98) to[bend left=22] (axis cs:1.45, 0.852);
  \draw[dashed, gray] (axis cs:1.5,0)     -- (axis cs:1.5, 0.8333);
  \draw[dashed, gray] (axis cs:0, 0.8333) -- (axis cs:1.5, 0.8333);
  \node[below=20pt, font=\small] at (axis cs:1.25, 0)
    {$E_{\mathrm{total}} = E_{\mathrm{PoUW}} = \tfrac{3}{2}$};
\end{axis}
\end{tikzpicture}
\caption{Supply and demand for \cref{ex:linear_demand} ($\gamma=3$, $R=\tfrac{1}{4}$), before and after the transition to PoUW. Following the economics convention, the price $P$ is on the vertical axis and the quantity $Q$ on the horizontal. \emph{(a) Before (PoW):} the useful resource is supplied perfectly elastically at unit cost, $P_S^{\mathrm{PoW}}(Q)=1$, so the equilibrium with demand $D$ lies at $A=(1,1)$; the shaded square (area~$1$) is the electricity spent on useful work, while PoW mining burns an additional $R=\tfrac{1}{4}$, for a total of $\tfrac{5}{4}$. \emph{(b) After (PoUW):} redistributing the mining reward as a per-unit subsidy $R/Q$ turns the supply curve into the rectangular hyperbola $P_S^{\mathrm{PoUW}}(Q)=1-R/Q$, which meets the axis $P=0$ at $Q=R=\tfrac{1}{4}$. The equilibrium moves along the demand curve from $A$ down and to the right to $B=(\tfrac{3}{2},\tfrac{5}{6})$, i.e., to consumer price $P^*=\tfrac{5}{6}$ and output $Q^*=\tfrac{3}{2}$; since every unit's cost is electricity, total consumption equals the shaded area $Q^*\times 1=\tfrac{3}{2}$. Thus electricity \emph{rises} from $\tfrac{5}{4}$ to $\tfrac{3}{2}$ even though PoW has been eliminated.}
\label{fig:example_demand_supply}
\end{figure}
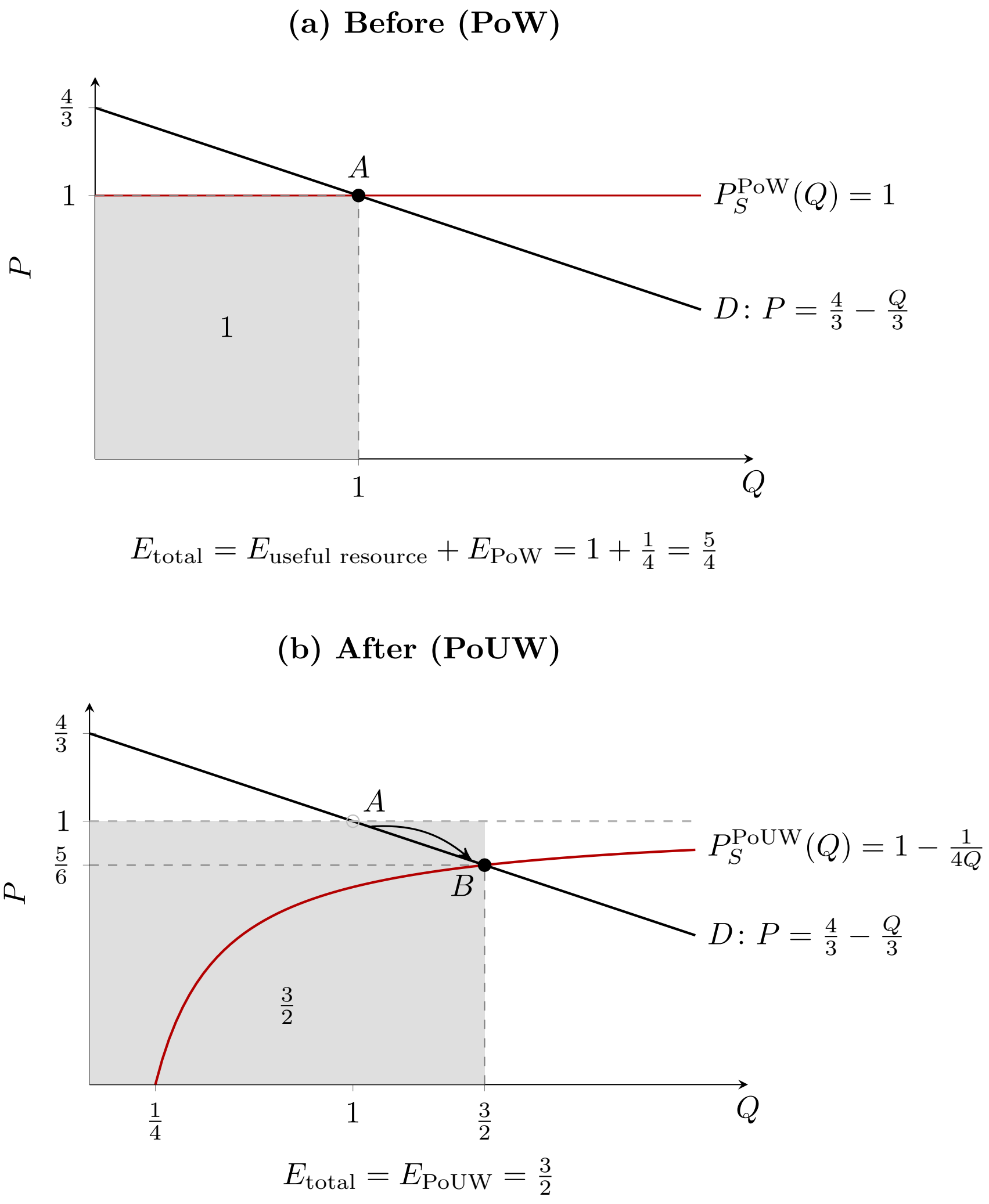 
 
\begin{remark}
    \Cref{prop:linear_demand_condition_for_increased_electricity_non_energy_costs} in \cref{sec:non_energy_costs} generalizes \cref{prop:linear_demand_condition_for_increased_electricity} to the setting where only a fraction of production costs are spent on electricity.
\end{remark}

\subsection{General Demand: Small-Reward Regime}
A natural regime to focus on is the one in which the original expenditure on the useful resource (e.g., MatMul), which is normalized to $\$1$, is much larger than the mining reward $R$, i.e., $R \ll 1$.

\begin{assumption}
\label{asm:smooth_demand}
    The demand curve $Q_D$ is continuous and nonincreasing (the law of demand), satisfies $Q_D(1)=1$, and is twice continuously differentiable near $P=1$.
\end{assumption}
We let $\gamma=-Q_D'(1)$.

\begin{proposition}
    Under \cref{asm:smooth_demand}, for all sufficiently small mining rewards $R>0$, total electricity consumption increases if $\gamma>1$.
    \label{prop:general_condition_for_increased_electricity}
\end{proposition}
\begin{proof}
Let $\gamma:=-Q_D'(1)$.
By a Taylor expansion around the pre-PoUW price $P=1$, evaluated at the post-PoUW price $P=1-\delta$, we have
\begin{equation}
\label{eq:general_demand_taylor}
    Q_D(1-\delta)=1+\gamma\delta+O(\delta^2).
\end{equation}
Thus, up to the second-order correction $O(\delta^2)$, the demand response is the same as for the linear demand curve of \cref{eq:linear_demand}.
The proof follows the linear-demand analysis, with the only change being that we carry along this second-order remainder.

Repeating the step leading to \cref{eq:linear_delta_condition}, with \cref{eq:general_demand_taylor} replacing \cref{eq:linear_demand}, gives
\[
    \delta
    =
    \frac{R}{1+\gamma\delta+O(\delta^2)}.
\]
This expansion is valid only near $P=1$, so we first check that $\delta$ is small when $R$ is. For $R<1\le Q_D(0)$, \cref{asm:positive_price} holds, so there is an equilibrium with a positive price. By \cref{eq:condition_on_delta}, $\delta=R/Q_D(1-\delta)>0$, so the post-PoUW price $1-\delta$ is below $1$, and since $Q_D$ is nonincreasing, $Q_D(1-\delta)\ge Q_D(1)=1$. Hence
\[
    0<\delta\le R.
\]
Thus $\delta=O(R)$; in particular, for all sufficiently small $R$, the price $1-\delta$ lies in the neighborhood of $P=1$ on which $Q_D$ is twice continuously differentiable, so the remainder in \cref{eq:general_demand_taylor} is at most a constant times $\delta^2$.
Substituting this back into the denominator gives
\[
    1+\gamma\delta+O(\delta^2)=1+O(R),
\]
and therefore
\[
    \delta=\frac{R}{1+O(R)}=R+O(R^2).
\]

In the linear case, \cref{eq:linear_post_pouw_quantity} shows that the post-PoUW electricity consumption is exactly $1+\gamma\delta$.
Here, by \cref{eq:general_demand_taylor}, it is
\[
    Q_D(1-\delta)
    =
    1+\gamma\delta+O(\delta^2)
    =
    1+\gamma R+O(R^2).
\]
The pre-PoUW analysis does not change from \cref{prop:linear_demand_condition_for_increased_electricity}: electricity consumption before the transition is $1+R$, consisting of one unit from the useful-resource market by the normalization $Q_D(1)=1$, plus $R$ from PoW mining.
Therefore, electricity consumption increases if and only if
\[
    Q_D(1-\delta)>1+R.
\]
Using the expansion above, the change in total electricity consumption is
\[
    Q_D(1-\delta)-(1+R)
    =(\gamma-1)R+O(R^2)
    =R\bigl(\gamma-1+O(R)\bigr).
\]
For a fixed demand curve with $\gamma>1$, the positive first-order term dominates the remainder for sufficiently small $R>0$, so electricity consumption increases.
Thus, for sufficiently small mining rewards, elastic demand at the pre-PoUW equilibrium leads to increased electricity consumption.
\end{proof} 
\begin{remark} 
    \Cref{prop:general_condition_for_increased_electricity_non_energy_costs} in \cref{sec:non_energy_costs} generalizes \cref{prop:general_condition_for_increased_electricity} to the setting where only a fraction of production costs are spent on electricity.
\end{remark}

\section{Discussion}
\label{sec:discussion}
The analysis above applies most directly when the useful work is a privately demanded market good. Two variants fall outside the scope of the main analysis.

\begin{itemize}
    \item \textbf{Public-good useful work.}
    Our analysis focuses on useful work that is a privately demanded market good, such as MatMul: users buy more of it when the effective consumer price falls. This demand response is what creates the rebound or backfire effect. The situation is different for PoUW schemes whose output is useful only as a public good. For example, Primecoin~\cite{Kin13} uses mining to search for prime-number structures, whose value is not naturally sold unit by unit to private consumers. In such a setting, there is no consumer price whose reduction induces additional private demand. With the mining reward held fixed, the transition from PoW to PoUW therefore redirects the protocol's existing mining budget from hashing to producing the public good, rather than expanding a private market for the useful resource. Thus the backfire mechanism identified above does not apply: electricity expenditure would not increase through induced demand.

    \item \textbf{PoUW overhead.}
    Our analysis assumes that producing a proof of useful work costs no more than the useful computation itself. In practice a proof carries some overhead: in the MatMul-based construction, for instance, the perturbation and hashing steps that make the result verifiable are extra work; the overhead factor is $1+o(1)$ in the matrix dimension~\cite{KW26}, so the overhead vanishes asymptotically, but it need not be negligible at realistic sizes. We measure it by the factor $\alpha\ge1$ by which a proof multiplies the compute of the bare useful computation; a factor of $2$, for example, means that providing the proof doubles the compute.

    A producer who turns its useful work into proofs of useful work pays this extra compute and, in return, earns a share of the mining reward, and producers take up proof production as long as it pays. What matters is how the reward compares with the extra compute the whole market would need. If the reward is too small to pay for the overhead of every producer, only some of them make proofs: they enter until the reward just covers their extra compute, while the others keep selling the useful resource at cost, so competition holds its price where it was. Consumers see no discount, the reward is burned on the overhead of the proofs just as it was burned on hashing, and electricity consumption is exactly what it was before the transition. Since the mining reward is typically small relative to the market for useful computation, even a modest overhead leads to this case. If instead the reward exceeds the overhead of the entire market, every producer makes proofs; the overhead absorbs the first part of the subsidy, the remainder reaches consumers as a discount, and the analysis of this paper applies. \Cref{sec:overhead} makes this precise, following Pass~\cite{Pas26} and his convention that compute spent on proofs secures the chain as well as hashing does. With the useful-work market normalized to size $1$ as in \cref{sec:model}, the first case is $\alpha\ge 1+R$ and the second is $\alpha<1+R$, where the backfire threshold of \cref{prop:linear_demand_condition_for_increased_electricity} becomes $\gamma>(1+R)/\alpha$: with overhead, backfire requires \emph{less} demand elasticity, because every unit of additional output now costs more electricity.
\end{itemize}

Our analysis concerns electricity consumption alone. The transition from PoW to PoUW also affects the security of the blockchain and social welfare. We close by discussing both, drawing on the equilibrium analysis of Pass~\cite{Pas26}.

\begin{itemize}
    \item \textbf{Security budget.}
    This work compares the implications for \emph{electricity consumption} of transitioning from PoW to PoUW while maintaining the mining security budget. One might argue that PoUW needs a smaller security budget than PoW, since the useful-work market already pays for much of the compute that secures the chain. In terms of raw compute, this is true: in Pass's Fortessia regime, for instance, duplex work replaces hashing and raises the amount of compute securing the chain for the same reward~\cite{Pas26}. This regime arises only when the PoUW overhead is non-negligible: it is the case $\alpha\ge 1+R$ of the PoUW-overhead item above. What matters to an attacker, however, is the economic cost of a majority attack, and an attacker who does useful work is paid for it at the market price. Pass shows that, once equilibrium prices are taken into account, this cost equals half the dollar value of the block reward in every regime, the same as under PoW~\cite{Pas26}. The same result refutes the opposite, na\"ive argument that PoUW is cheaper to attack because the attacker is ``paid to attack.'' The useful-work market thus does not substitute for the block reward, and cutting the dollar value of the reward lowers the cost of an attack proportionally, under PoUW just as under PoW.

    \item \textbf{Welfare analysis.}
    Pass~\cite[Section~4.3]{Pas26} studies how the transition from PoW to PoUW affects social welfare.
     Social welfare is the value of the useful resource to its consumers, plus the value of the blockchain itself, minus the cost of all compute; prices are transfers between consumers and producers and do not count. 
    When the dollar reward is held fixed, as in our model, he finds that PoUW (his Duplexia) strictly dominates PoW: PoW burns the reward on hashing, whereas PoUW spends the same reward and also produces useful output that consumers value. His preferred comparison fixes the token--inference ratio $\theta$ instead. When demand is elastic, the dollar value of the reward under PoUW then grows with spending on useful work, and so do the subsidy and the losses on the additional units, while the saving from avoided hashing stays at the smaller PoW reward; this is why that comparison can turn against PoUW, as it does under isoelastic demand once $\theta>1/(\varepsilon-1)$.

    Social welfare is easy to evaluate in our model, which corresponds to Pass's with no overhead and a fixed dollar reward. Like Pass, we hold the value of the blockchain fixed, which is natural here, since the cost of a majority attack does not change (see the previous item). Under PoUW, the reward reaches consumers as a discount, and the only loss is on the $Q^*-1$ additional units, each of which costs $1$ to produce but is worth less than that to its buyer. With the linear demand of \cref{eq:linear_demand}, this loss is the triangle $\gamma\delta^2/2$, and since $R=\delta+\gamma\delta^2$ by \cref{eq:linear_delta_condition}, the change in social welfare is
    \[
        \Delta W = R-\frac{\gamma\delta^2}{2} = \frac{R+\delta}{2} > 0,
    \]
    which is exactly the gain in consumer surplus. This gain is positive for every $\gamma$, whereas electricity consumption grows if and only if $\gamma>1+R$ (\cref{prop:linear_demand_condition_for_increased_electricity}). In \cref{ex:linear_demand}, welfare rises by $5/24$, more than $80\%$ of the reward, while electricity consumption rises by $20\%$. This measure does not account for the external costs of electricity, such as emissions: it values electricity at its price. If each dollar of electricity causes $\tau$ dollars of external damage, a transition that backfires lowers welfare once $\tau$ exceeds $\Delta W$ divided by the increase in electricity consumption, a ratio that is $5/6$ in \cref{ex:linear_demand} and tends to $1/(\gamma-1)$ as $R\to 0$. When demand is sufficiently elastic, PoUW thus benefits the users of the useful resource while increasing electricity consumption.

\end{itemize}
\paragraph{AI Disclosure.} The main ideas, the model, and the derivations of the main results are our own. We used Claude (Anthropic) and ChatGPT (OpenAI) for proofreading and routine editorial work, to produce the graphs, to help with the literature review and with some of the derivations in \cref{sec:non_energy_costs,sec:overhead}, all of which we checked. The authors take full responsibility for the entire content of this paper.

\ifnum\anonymous=0
    \ifnum\lipics=0
        \ifnum\masterthesis=0
            \subsection*{Acknowledgments}
    We thank Ittay Eyal and Shai Wyborski for fruitful discussions and comments. This research was supported by the Israel Science Foundation (grant No.\ 2527/24) and by the Council for Higher Education through the Research Center for the Theory of Quantum Computing.

            \BeforeBeginEnvironment{wrapfigure}{\setlength{\intextsep}{0pt}}
            \begin{wrapfigure}{r}{100px}
                \includegraphics[width=100px]{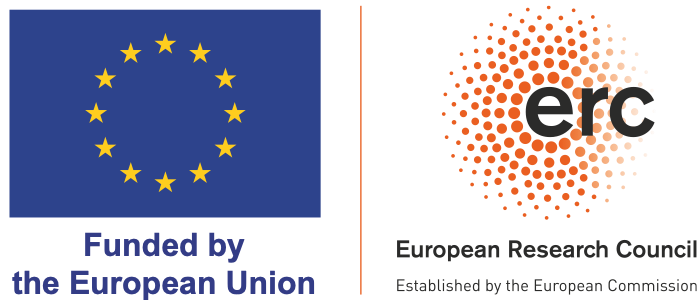}
            \end{wrapfigure}
            This work was funded by the European Union (ERC-2022-COG, ACQUA, 101087742). Views and opinions expressed are however those of the authors only and do not necessarily reflect those of the European Union or the European Research Council Executive Agency. Neither the European Union nor the granting authority can be held responsible for them.
            
        \fi
    \fi
\fi
\ifnum\sigconf=1
    \bibliographystyle{ACM-Reference-Format}
\else
    \ifnum\cryptology=1
        \bibliographystyle{abbrv}
    \else
        \ifnum\lipics=1
            \bibliographystyle{plainurl}
        \else
            \ifnum\llncs=1
                \bibliographystyle{splncs04}
            \else 
                \bibliographystyle{alphaabbrurldoieprint}
            \fi
        \fi
    \fi
\fi
\ifnum\masterthesis=0
    \ifnum\smallbib=1
        {\footnotesize \bibliography{main} }
    \else 
        \bibliography{main}
    \fi
\fi

\crefname{appendix}{Appendix}{Appendices}
\Crefname{appendix}{Appendix}{Appendices}
\appendix
\crefalias{section}{appendix}
\crefalias{subsection}{appendix}
\counterwithin{proposition}{section}
\counterwithin{lemma}{section}
\counterwithin{corollary}{section}
\ifnum\shownomenclature=1
    \printnomenclature[1in]
\fi
\section{Accounting for Non-Energy Costs}
\label{sec:non_energy_costs}

We now extend the model to the case where only a fraction of the total cost is spent on electricity. We work in an amortized-cost model, so there is no distinction between fixed costs and per-unit costs. Thus, the cost of producing one unit of the useful resource is still normalized to $1$, but only a fraction of this cost is spent on electricity.

Let $e \in [0,1]$ denote the fraction of the PoW mining cost that is spent on electricity, and let $e' \in (0,1]$ denote the fraction of the cost of the useful resource that is spent on electricity. Thus, prior to the transition to PoUW, the total electricity expenditure is
\[
e' + eR,
\]
since one unit of the useful resource is produced at price $1$, and the mining reward $R$ is fully dissipated in equilibrium. After the transition to PoUW, the total amount of useful work produced is $Q_D(1-\delta)$, and so the total electricity expenditure is
\[
e'Q_D(1-\delta).
\]
 
Importantly, the equilibrium condition determining $\delta$ does not change. Indeed, the total cost of producing $Q_D(1-\delta)$ units is still $Q_D(1-\delta)$, while the total revenue is the payment by the users plus the subsidy, namely
\[
(1-\delta)Q_D(1-\delta) + R.
\]  
Equating cost and revenue gives
\[
Q_D(1-\delta) = (1-\delta)Q_D(1-\delta) + R,
\]
which is equivalent to \cref{eq:condition_on_delta}:
\[
\delta = \frac{R}{Q_D(1-\delta)}.
\] 

Thus, electricity expenditure increases under the transition to PoUW if and only if
\[
e'Q_D(1-\delta) > e' + eR,
\]
or, equivalently,
\begin{equation}
\label{eq:non_energy_backfire_condition}
Q_D(1-\delta) > 1 + \frac{e}{e'}R.
\end{equation}

We now extend \cref{prop:linear_demand_condition_for_increased_electricity,prop:general_condition_for_increased_electricity} under this more general accounting.

\begin{proposition}
In the linear demand model with $0<R<1+\gamma$, the total electricity consumption grows when transitioning to PoUW if and only if
\[
\gamma > \frac{e}{e'} + \left(\frac{e}{e'}\right)^2 R.
\]
\label{prop:linear_demand_condition_for_increased_electricity_non_energy_costs}
\end{proposition}
 
\begin{proof}
The equilibrium condition \cref{eq:condition_on_delta} becomes
\[
\delta \stackrel{\text{by \cref{eq:linear_demand}}}{=} \frac{R}{1+\gamma\delta}.
\]
The positive solution is
\[
\delta = \frac{-1+\sqrt{1+4\gamma R}}{2\gamma}.
\]
Hence
\[
Q_D(1-\delta) = 1+\gamma\delta = \frac{1+\sqrt{1+4\gamma R}}{2}.
\]

Substituting this expression from \cref{eq:linear_post_pouw_quantity} into \cref{eq:non_energy_backfire_condition} shows that electricity expenditure increases if and only if
\[
\frac{1+\sqrt{1+4\gamma R}}{2} > 1 + \frac{e}{e'}R,
\]
which is equivalent to
\[
\sqrt{1+4\gamma R} > 1 + 2\frac{e}{e'}R.
\]
Squaring both sides gives
\[
1+4\gamma R > 1 + 4\frac{e}{e'}R + 4\left(\frac{e}{e'}\right)^2 R^2,
\]
which is equivalent to
\[
\gamma > \frac{e}{e'} + \left(\frac{e}{e'}\right)^2 R.
\]
\end{proof}

When $e=e'$, this reduces to the condition $\gamma > 1+R$ of \cref{prop:linear_demand_condition_for_increased_electricity}, as expected: electricity is then the same fraction $e$ of every cost, so electricity consumption, both before and after the transition, is $e$ times its value in the original model, which is the case $e=e'=1$.

Of course, the condition $e=e'$ may not hold: a reasonable assumption is that $e>e'$, namely, that PoW mining is more electricity-intensive than the useful resource produced by the PoUW. Indeed, MatMul may involve a larger fraction of non-energy costs, most notably, a larger fraction of the cost seems to be spent on GPU hardware (compared to electricity). In this case, the threshold for an increase in electricity consumption becomes larger than in the model where $e=e'$, and therefore, this effect is favorable for PoUW from an energy-consumption perspective.

For illustration, suppose that electricity accounts for $80\%$ of the cost of PoW mining but only $40\%$ of the cost of the useful resource, so that $e=0.8$, $e'=0.4$, and $e/e'=2$. By \cref{prop:linear_demand_condition_for_increased_electricity_non_energy_costs}, electricity consumption then increases if and only if $\gamma>2+4R$, compared with $\gamma>1+R$ when all costs are spent on electricity. Moderately elastic demand, say $\gamma=1.5$, thus leads to backfire in the original model whenever $R<\tfrac12$, but never here.

We keep \cref{asm:smooth_demand} on $Q_D$, and $\gamma=-Q_D'(1)$. By $R\ll 1$ we mean that the proposition holds for all sufficiently small mining rewards $R>0$.

\begin{proposition}
Under \cref{asm:smooth_demand}, when $R \ll 1$, the electricity consumption increases if
\[
-Q_D'(1) > \frac{e}{e'}.
\]
\label{prop:general_condition_for_increased_electricity_non_energy_costs}
\end{proposition}

\begin{proof}
Let
\[
\gamma := -Q_D'(1).
\]
The Taylor expansion in \cref{eq:general_demand_taylor} is valid only near $P=1$, so we first check that $\delta$ is small. Exactly as in the proof of \cref{prop:general_condition_for_increased_electricity}, the equilibrium condition \cref{eq:condition_on_delta} and the law of demand give $0<\delta\le R$. Hence, for $R\ll 1$, the expansion applies, and $Q_D(1-\delta)=1+O(\delta)$, so it follows that
\begin{equation}
\label{eq:non_energy_discount_expansion}
\delta = R + O(R^2).
\end{equation}
Consequently,
\begin{equation}
\label{eq:non_energy_small_reward_quantity}
\begin{aligned}
Q_D(1-\delta)
&\stackrel{\text{by \cref{eq:general_demand_taylor}}}{=} 1 + \gamma\delta + O(\delta^2)\\
&\stackrel{\text{by \cref{eq:non_energy_discount_expansion}}}{=} 1 + \gamma R + O(R^2).
\end{aligned}
\end{equation}

Substituting \cref{eq:non_energy_small_reward_quantity} into \cref{eq:non_energy_backfire_condition} shows that electricity expenditure increases if and only if
\[
1+\gamma R + O(R^2) > 1 + \frac{e}{e'}R,
\]
which is equivalent to
\[
\gamma > \frac{e}{e'} + O(R),
\]
where the $O(R)$ term is the negative of the remainder $O(R^2)$ in \cref{eq:non_energy_small_reward_quantity} divided by $R$, a fixed function of $R$ that tends to $0$ as $R\to 0$. Hence, if $\gamma>e/e'$, this inequality holds for all sufficiently small $R>0$. Recalling that $\gamma=-Q_D'(1)$ completes the proof.
\end{proof}

This shows that once non-energy costs are taken into account, the threshold is no longer universal. In the original model, the threshold was determined by whether demand is elastic at the operating point. In the present model, the relevant threshold is scaled by the ratio $e/e'$. In particular, if PoW is more electricity-intensive than the useful resource, then the transition to PoUW requires a larger demand response in order to increase total electricity consumption.

\section{Accounting for PoUW Overhead}
\label{sec:overhead}

Our model assumes that producing a proof of useful work costs exactly as much as producing the useful work itself. In practice, a proof might carry some overhead. In the MatMul-based construction of~\cite{KW26}, for instance, a provider who only wants the product can skip the perturbation and hashing steps that make the result publicly verifiable; the resulting overhead factor is $1+o(1)$ in the matrix dimension, so the overhead vanishes for large matrices, but it need not be negligible in practice. This appendix takes a possible overhead into account. It was written after the work of Pass~\cite{Pas26} appeared, in reaction to it, and it is the only part of this paper that postdates his work. We follow Pass's conventions: one unit of compute spent on proofs of useful work yields $1/\alpha$ units of the useful resource, where $\alpha\ge 1$ is the overhead parameter, and one unit of security, as a unit of hashing does (Pass's mining-side overhead, which we denote by $\gamma_{\mathrm{Pass}}$, is $1$). Under this assumption, proofs are always more efficient than mining and useful work done separately, so only two of his three regimes arise, and we borrow his names for them. Throughout, $R$ denotes the mining reward, and the pre-PoUW useful-work market is normalized to size $1$, as in \cref{sec:model}, so that the pre-PoUW electricity consumption is $1+R$.

Informally, our results are as follows.
\begin{itemize}
    \item \textbf{Large overhead} ($\alpha\ge 1+R$, which Pass calls \emph{Fortessia}). Producing proofs is not worthwhile for most of the market. Only a small fraction of producers switch to producing proofs, pure useful work and proof-producing useful work coexist, and the entire mining reward is burned on the overhead of the proofs. The price and quantity of the useful resource are unchanged, and so is the total electricity consumption: it is exactly $1+R$, as before the transition. In this regime, PoUW behaves like PoW as far as electricity is concerned; the reward is merely spent on overhead instead of hashing.
    \item \textbf{Small overhead} ($\alpha<1+R$, which Pass calls \emph{Duplexia}). The entire market switches to producing proofs. The first $\alpha-1$ dollars of the per-unit subsidy are absorbed by the overhead, and only the remainder is passed through to consumers as a discount. The backfire condition of \cref{prop:linear_demand_condition_for_increased_electricity} changes from $\gamma>1+R$ to
    \[
        \gamma>\frac{1+R}{\alpha}.
    \]
    Perhaps surprisingly, the threshold is \emph{lower} than without overhead: the overhead makes backfire easier. The discount, and hence the demand expansion, is smaller, but every unit of output now costs $\alpha>1$ units of electricity, and the second effect wins.
\end{itemize}
Combining the two regimes, we find that electricity consumption grows if and only if $(1+R)/\gamma<\alpha<1+R$ (\cref{cor:overhead_window}). In particular, the transition backfires for some overhead level if and only if $\gamma>1$, i.e., if and only if demand is elastic at the pre-PoUW equilibrium, without the correction term $R$ of \cref{prop:linear_demand_condition_for_increased_electricity}.

We now turn to the formal treatment. Pass~\cite{Pas26} allows for an overhead as follows. Using our notation, his model allocates compute to three activities, each costing one unit per operation: \emph{pure mining} $M$, which produces one unit of security and nothing useful; \emph{pure useful work} $U$, which produces one unit of the useful resource and no security; and \emph{duplex} work $H$, which produces both, but at reduced efficiency: one duplex operation yields $1/\alpha$ units of the useful resource and $1/\gamma_{\mathrm{Pass}}$ units of security, where $\alpha,\gamma_{\mathrm{Pass}}\ge 1$ are technological overhead parameters.\footnote{Pass denotes the mining-side overhead by $\gamma$. Since $\gamma$ denotes minus the slope of the demand curve $Q_D(P)$ throughout this paper (see \cref{eq:linear_demand}), we write $\gamma_{\mathrm{Pass}}$ for his parameter. It plays no role below, since we set it to $1$.} In the MatMul example above, $\alpha>1$ captures the cost of the perturbation and hashing steps.

We incorporate the overhead $\alpha$ into our analysis using the simplest version of Pass's model. We make the following simplifying assumptions:
\begin{enumerate}
    \item \textbf{Fixed dollar reward.} As in the rest of this paper, the total mining reward $R$ (in dollars per unit time) is held fixed across the transition. Pass instead holds fixed the \emph{token-inference ratio} $\theta$, the ratio of the dollar value of block rewards to the dollar value of useful work transacted. At our pre-PoUW equilibrium, useful-work spending is $1$ by normalization, so $\theta=R$ there; Pass's thresholds in $\theta$ thus become thresholds in $R$ below.
    \item \textbf{No mining-side overhead}, i.e., $\gamma_{\mathrm{Pass}}=1$: a duplex operation secures the chain as well as a pure hashing operation does. Only the useful-work overhead $\alpha\ge 1$ remains.
    \item \textbf{Linear demand}, $Q_D(P)=1+\gamma(1-P)$ as in \cref{eq:linear_demand}, with $\gamma>0$.
    \item \textbf{Electricity-only costs}, as in \cref{sec:model} (so $e=e'=1$ in the notation of \cref{sec:non_energy_costs}).
\end{enumerate}

\paragraph{Setup.}
Let $M,U,H\ge 0$ denote the amounts of compute (per unit time) devoted to pure mining, pure useful work, and duplex (hybrid) work, respectively. Total security is $S=M+H$ (recall $\gamma_{\mathrm{Pass}}=1$), and total useful output is
\begin{equation}
\label{eq:overhead_output}
    Q = U + \frac{H}{\alpha}.
\end{equation}
Let $P$ denote the consumer price of one unit of the useful resource. The mining reward $R$ is shared among contributors in proportion to their security, so one unit of pure mining earns $R/S$, and one duplex operation earns $R/S$ from mining plus $P/\alpha$ from selling its useful output. Since each operation costs $1$, the per-operation profits are
\begin{equation}
\label{eq:overhead_profits}
    \pi_M = \frac{R}{S}-1,
    \qquad
    \pi_U = P-1,
    \qquad
    \pi_H = \frac{R}{S}+\frac{P}{\alpha}-1.
\end{equation}
With free entry, competitive equilibrium requires $\pi_X\le 0$ for every activity $X\in\{M,U,H\}$, with equality for every activity that is active ($X>0$)~\cite{Pas26}. Total electricity consumption is the total compute, $M+U+H$.

Before the transition, duplex work is unavailable ($H=0$). Then $\pi_M=0$ gives $M=S=R$ (the reward is fully dissipated in hashing) and $\pi_U=0$ gives $P=1$, hence $U=Q_D(1)=1$. Total electricity consumption is $1+R$, as in \cref{sec:model}.

After the transition, pure mining is never active. Indeed, \cref{eq:overhead_profits} gives $\pi_H-\pi_M=P/\alpha>0$ whenever $P>0$, so if $M>0$, then $\pi_M=0$ would force $\pi_H>0$, contradicting the equilibrium condition. Since $M=0$, we have $S=H$, and $\pi_H=0$ becomes
\begin{equation}
\label{eq:duplex_zero_profit}
    \frac{R}{H}+\frac{P}{\alpha}=1.
\end{equation}
Two cases remain, depending on whether pure useful work $U$ is active. Following Pass, we call them \emph{Fortessia} (duplex coexists with pure useful work) and \emph{Duplexia} (all useful work is duplex).

\subsection{Large Overhead: Electricity Is Unchanged}

\begin{lemma}[Fortessia]
\label{lem:overhead_fortessia}
If $\alpha\ge 1+R$, then in equilibrium $P=1$, $Q=1$,
\[
    H=\frac{\alpha R}{\alpha-1},
    \qquad
    U=1-\frac{R}{\alpha-1}\ge 0,
\]
and the total electricity consumption is $H+U=1+R$, exactly as before the transition.
\end{lemma}
\begin{proof}
Suppose first that $U>0$, so $\pi_U=0$ and $P=1$. Then \cref{eq:duplex_zero_profit} gives $R/H=1-1/\alpha$, i.e., $H=\alpha R/(\alpha-1)$ (note that $\alpha>1$ since $\alpha\ge 1+R>1$). Demand at $P=1$ is $Q=Q_D(1)=1$, so \cref{eq:overhead_output} gives $U=Q-H/\alpha=1-R/(\alpha-1)$, which is nonnegative exactly when $\alpha-1\ge R$, i.e., $\alpha\ge 1+R$. Total electricity is
\[
    H+U=\frac{\alpha R}{\alpha-1}+1-\frac{R}{\alpha-1}=1+R.
\]
It remains to treat the case $U=0$ when $\alpha\ge 1+R$. If $U=0$, then $Q=H/\alpha$ and \cref{eq:duplex_zero_profit} gives $P=\alpha-R/Q$. The equilibrium condition $\pi_U\le 0$ requires $P\le 1$, i.e., $Q\le R/(\alpha-1)\le 1$. But $P\le 1$ also implies $Q=Q_D(P)\ge Q_D(1)=1$. Hence $Q=1$ and $P=1$, which is the same equilibrium as above (with $U=0$ only in the boundary case $\alpha=1+R$).
\end{proof}

In this regime, the subsidy does not reach the consumers at all: the price and quantity of the useful resource are unchanged, and the useful-work market simply absorbs the mining reward by running $H=\alpha R/(\alpha-1)$ operations of duplex work, of which $H(1-1/\alpha)=R$ is pure overhead. In other words, the $R$ dollars' worth of electricity that used to be burned on hashing is now burned on the overhead of the proofs, and total electricity consumption is exactly as before. This is the large-overhead regime described informally in \cref{sec:discussion}. Note also that the fraction of the useful-work market producing proofs is $H/(H+U)=\alpha R/((\alpha-1)(1+R))$, which is small when $R\ll \alpha-1$.

\subsection{Small Overhead: A Modified Backfire Threshold}

We now derive the analogue of \cref{prop:linear_demand_condition_for_increased_electricity} in the presence of overhead. Recall that without overhead, electricity consumption grows if and only if $\gamma>1+R$. The next proposition shows that with overhead $\alpha<1+R$, the threshold becomes $\gamma>(1+R)/\alpha$: it is \emph{lower} than without overhead, so backfire becomes \emph{easier}. As the proof below shows, in this regime all useful work is produced by duplex operations. Without overhead, a producer breaks even when $P+R/Q=1$: the consumer price plus the per-unit subsidy covers the unit cost of $1$. With overhead, each unit of the useful resource requires $\alpha$ duplex operations and thus costs $\alpha$, while the per-unit subsidy is still $R/Q$ (all of the reward now goes to the duplex producers). Hence the break-even condition becomes $P+R/Q=\alpha$, and the supply curve in \cref{eq:supply_curve} shifts upward by $\alpha-1$:
\begin{equation}
\label{eq:supply_curve_overhead}
    P_S^{\mathrm{PoUW},\alpha}(Q)=\alpha-\frac{R}{Q}.
\end{equation}
Equivalently, in terms of the discount $\delta=1-P^*$ relative to the pre-PoUW price, \cref{eq:condition_on_delta} becomes
\[
    \delta=\frac{R}{Q_D(1-\delta)}-(\alpha-1).
\]
The first $\alpha-1$ dollars of the per-unit subsidy are absorbed by the overhead, and only the remainder is passed through to consumers. Thus the overhead changes the analysis of \cref{prop:linear_demand_condition_for_increased_electricity} in two ways. First, the demand expansion is smaller, because the effective subsidy is reduced from $R$ to $R-(\alpha-1)Q^*$; this works against backfire. Second, each unit produced now consumes $\alpha$ units of electricity rather than $1$, so the post-PoUW consumption is $\alpha Q^*$ rather than $Q^*$; this works in favor of backfire. As the proposition below shows, the second effect wins, and the net result is that the threshold is divided by $\alpha$.

\begin{proposition}
\label{prop:overhead_linear_threshold}
Let $\alpha<1+R$. In the linear demand model with overhead $\alpha$, the total electricity consumption grows when transitioning to PoUW if and only if
\[
    \gamma>\frac{1+R}{\alpha}.
\]
\end{proposition}
\begin{proof}
Free entry into pure useful work and \cref{eq:overhead_profits} imply $P^*\le 1$. If $P^*=1$, demand is one, so total production costs at most $\alpha$, even if all output is produced by duplex operations. But total producer revenue is $1+R>\alpha$, contradicting zero profit. Thus $P^*<1$, making pure useful work unprofitable, so $U=0$.

Define the equilibrium discount by
\begin{equation}
\label{eq:overhead_positive_discount}
    \delta:=1-P^*>0.
\end{equation}
Since \cref{eq:overhead_output} gives $H=\alpha Q^*$, we obtain
\begin{equation}
\label{eq:overhead_equilibrium}
    R\stackrel{\text{by \cref{eq:duplex_zero_profit}}}{=}(\alpha-1+\delta)Q^*,
    \qquad Q^*\stackrel{\text{by \cref{eq:linear_demand}}}{=}1+\gamma\delta.
\end{equation}
Adding $Q^*$ to both sides of the first equality in \cref{eq:overhead_equilibrium} and substituting $Q^*=1+\gamma\delta$ on the left gives
\begin{equation}
\label{eq:overhead_quantity_identity}
    (\alpha+\delta)Q^*\stackrel{\text{by \cref{eq:overhead_equilibrium}}}{=}1+R+\gamma\delta.
\end{equation}
The post-PoUW electricity consumption is $H=\alpha Q^*$, so its change from the pre-PoUW level is
\[
    \alpha Q^*-(1+R).
\]
We can rewrite this difference as
\[
    \begin{aligned}
        \alpha Q^*-(1+R)
        &\stackrel{\text{by \cref{eq:overhead_quantity_identity}}}{=}\frac{\alpha(1+R+\gamma\delta)}{\alpha+\delta}-(1+R)\\
        &=\frac{\alpha(1+R+\gamma\delta)-(1+R)(\alpha+\delta)}{\alpha+\delta}\\
        &=\frac{\delta}{\alpha+\delta}\bigl(\alpha\gamma-(1+R)\bigr).
    \end{aligned}
\]
The last equality cancels the two $\alpha(1+R)$ terms and factors out $\delta$. Since $\delta>0$ by \cref{eq:overhead_positive_discount} and $\alpha\ge 1$, the prefactor $\delta/(\alpha+\delta)$ is positive. Therefore, electricity consumption increases if and only if $\gamma>(1+R)/\alpha$.
\end{proof}


Combining \cref{lem:overhead_fortessia,prop:overhead_linear_threshold} gives a complete picture of how the overhead affects our main result.

\begin{corollary}
\label{cor:overhead_window}
Fix $R>0$ and $\gamma>0$. In the linear demand model with overhead $\alpha\ge 1$, the total electricity consumption grows when transitioning to PoUW if and only if
\[
    \frac{1+R}{\gamma}<\alpha<1+R.
\]
In particular, electricity consumption grows for some overhead level $\alpha\ge 1$ if and only if $\gamma>1$, i.e., if and only if demand is elastic at the pre-PoUW equilibrium.
\end{corollary}
\begin{proof}
For $\alpha\ge 1+R$, electricity consumption is unchanged by \cref{lem:overhead_fortessia}. For $\alpha<1+R$, it grows if and only if $\gamma>(1+R)/\alpha$ by \cref{prop:overhead_linear_threshold}, i.e., if and only if $\alpha>(1+R)/\gamma$. The interval $\bigl((1+R)/\gamma,\,1+R\bigr)$ is non-empty if and only if $\gamma>1$.
\end{proof}

\Cref{cor:overhead_window} refines the conclusions of \cref{sec:main_results} in three ways.
\begin{itemize}
    \item If $\gamma>1+R$, so that electricity consumption grows already without overhead, then it grows for every overhead $\alpha<1+R$; only an overhead of at least $1+R$ neutralizes the effect, by bringing consumption back to its pre-PoUW level (never below it).
    \item If $1<\gamma\le 1+R$, then without overhead the transition does not increase electricity consumption (it reduces it if $\gamma<1+R$ and leaves it unchanged if $\gamma=1+R$), but a moderate overhead $\alpha\in\bigl((1+R)/\gamma,\,1+R\bigr)$ \emph{creates} backfire: part of the subsidy is burned as overhead, while the remaining discount still expands output, and each additional unit of output now costs $\alpha>1$ units of electricity.
    \item If $\gamma\le 1$, then electricity consumption never grows, whatever the overhead. Thus, once overhead is taken into account, the condition that demand is elastic at the pre-PoUW equilibrium is exactly the condition under which backfire is possible for some overhead level, without the correction term $R$ in \cref{prop:linear_demand_condition_for_increased_electricity}.
\end{itemize}

\begin{example}[\Cref{ex:linear_demand} with overhead]
\label{ex:overhead}
Take $\gamma=3$ and $R=\tfrac14$ as in \cref{ex:linear_demand}, so that $1+R=\tfrac54$ and the pre-PoUW electricity consumption is $\tfrac54$.
For $\alpha\ge\tfrac54$, the economy is in Fortessia and consumption stays at $\tfrac54$. For $\alpha=1.1$, \cref{eq:overhead_equilibrium} gives $(1+3\delta)(0.1+\delta)=\tfrac14$, whose positive solution is $\delta\approx 0.095$, so the consumer price falls to $P^*\approx 0.905$, output rises to $Q^*\approx 1.284$, and electricity consumption rises to $\alpha Q^*\approx 1.41$, an increase of about $13\%$ (compared with $20\%$ at $\alpha=1$). In this example, consumption decreases monotonically in $\alpha$ from $\tfrac32$ at $\alpha=1$ to $\tfrac54$ at $\alpha=\tfrac54$ (for instance, $\alpha=1.2$ gives approximately $1.31$), so here the overhead dampens the backfire without eliminating it for any $\alpha<\tfrac54$, as predicted by the first bullet above, since $\gamma=3>\tfrac54$.

The second bullet is illustrated by $\gamma=\tfrac43$ and $R=0.42$, so that $1+R=1.42$. Without overhead, $\gamma<1+R$ and \cref{prop:linear_demand_condition_for_increased_electricity} predicts a decrease: indeed, $\delta=0.3$ and consumption falls from $1.42$ to $1+\gamma\delta=1.4$. With overhead $\alpha=1.2\in\bigl(\tfrac{1.42}{4/3},\,1.42\bigr)=(1.065,1.42)$, \cref{prop:overhead_linear_threshold} predicts an increase: $\delta=0.15$, $Q^*=1.2$, and consumption becomes $\alpha Q^*=1.44>1.42$.
\end{example}

\ifnum\masterthesis=1
    \ifnum\smallbib=1
        {\footnotesize \bibliography{main} }
    \else 
        \bibliography{main}

\begin{thebibliography}{dVGKS22}

\bibitem[BCLM25]{BCLM25}
J.~Brand, J.~C. Castillo, C.~Lohani, and L.~Musolff.
\newblock
  \href{https://chinmaylohani.com/files/Reliability_Draft.pdf}{Reliability and
  Pricing in Cloud Computing}.
\newblock Working paper, 2025.
\newblock Retrieved July 9, 2026.

\bibitem[BOKW25]{BKW25}
Y.~Bar-On, I.~Komargodski, and O.~Weinstein.
\newblock \href{https://arxiv.org/abs/2505.21685}{Proof of Work With External
  Utilities}, 2025,  arXiv: \href{https://arxiv.org/abs/2505.21685}{{\ttfamily
  2505.21685}}.

\bibitem[Bor15]{Bor15}
S.~Borenstein.
\newblock \href{https://doi.org/10.5547/01956574.36.1.1}{A Microeconomic
  Framework for Evaluating Energy Efficiency Rebound and Some Implications}.
\newblock {\em The Energy Journal}, 36(1):1--22, 2015.

\bibitem[BRSV17]{BRS+17}
M.~Ball, A.~Rosen, M.~Sabin, and P.~N. Vasudevan.
\newblock Proofs of Useful Work, 2017, Cryptology ePrint Archive: {{\ttfamily
  \href{https://eprint.iacr.org/2017/203.pdf}{Report 2017/203}}}.

\bibitem[{Cry}22]{CCRI22}
{Crypto Carbon Ratings Institute}.
\newblock \href{https://carbon-ratings.com/dl/eth-report-2022}{The Merge --
  Implications on the Electricity Consumption and Carbon Footprint of the
  {Ethereum} Network}.
\newblock Technical report, Crypto Carbon Ratings Institute ({CCRI}), September
  2022.
\newblock Retrieved October 23, 2026.

\bibitem[dV18]{Vri18}
A.~de~Vries.
\newblock \href{https://doi.org/10.1016/j.joule.2018.04.016}{Bitcoin's Growing
  Energy Problem}.
\newblock {\em Joule}, 2(5):801--805, May 2018.

\bibitem[dVGKS22]{VGKS22}
A.~de~Vries, U.~Gallersd{\"o}rfer, L.~Klaa{\ss}en, and C.~Stoll.
\newblock \href{https://doi.org/10.1016/j.joule.2022.02.005}{Revisiting
  Bitcoin's Carbon Footprint}.
\newblock {\em Joule}, 6(3):498--502, March 2022.

\bibitem[{Eth}22]{Eth22}
{Ethereum Foundation}.
\newblock
  \href{https://ethereum.org/developers/docs/consensus-mechanisms/pos/}{Proof-of-stake
  ({PoS})}.
\newblock ethereum.org, 2022.
\newblock Retrieved October 23, 2026.

\bibitem[FKKP19]{FKKP19}
A.~Fiat, A.~R. Karlin, E.~Koutsoupias, and C.~H. Papadimitriou.
\newblock \href{https://doi.org/10.1145/3328526.3329630}{Energy Equilibria in
  Proof-of-Work Mining}.
\newblock In A.~R. Karlin, N.~Immorlica, and R.~Johari, editors, {\em
  Proceedings of the 2019 {ACM} Conference on Economics and Computation, {EC}
  2019, Phoenix, AZ, USA, June 24-28, 2019}, pages 489--502. {ACM}, 2019.

\bibitem[Gal15]{Gal15}
R.~Galvin.
\newblock \href{https://doi.org/10.1016/j.ecolecon.2015.08.020}{The
  {ICT}/electronics question: Structural change and the rebound effect}.
\newblock {\em Ecological Economics}, 120:23--31, 2015.

\bibitem[GGD00]{Gre00}
L.~A. Greening, D.~L. Greene, and C.~Difiglio.
\newblock \href{https://doi.org/10.1016/S0301-4215(00)00021-5}{Energy
  efficiency and consumption --- the rebound effect --- a survey}.
\newblock {\em Energy Policy}, 28(6--7):389--401, 2000.

\bibitem[GKS20]{GKS20}
U.~Gallersd{\"o}rfer, L.~Klaa{\ss}en, and C.~Stoll.
\newblock \href{https://doi.org/10.1016/j.joule.2020.07.013}{Energy Consumption
  of Cryptocurrencies Beyond Bitcoin}.
\newblock {\em Joule}, 4(9):1843--1846, 2020.

\bibitem[Jev66]{Jev66}
W.~S. Jevons.
\newblock {\em The Coal Question: An Inquiry Concerning the Progress of the
  Nation, and the Probable Exhaustion of Our Coal-Mines}.
\newblock Macmillan, London, 2nd edition, 1866.

\bibitem[JLZ16]{JLZ16}
Q.~Jiang, Y.~C. Lee, and A.~Y. Zomaya.
\newblock \href{https://doi.org/10.1109/MCC.2016.14}{Price Elasticity in the
  Enterprise Computing Resource Market}.
\newblock {\em {IEEE} Cloud Comput.}, 3(1):24--31, 2016.

\bibitem[JPW26]{JPW26}
C.~Jin, S.~Peng, and P.~Wang.
\newblock \href{https://doi.org/10.2139/ssrn.4573326}{Firm Inertia in the
  Public Cloud}.
\newblock SSRN Working Paper 4573326, 2026.

\bibitem[Kin13]{Kin13}
S.~King.
\newblock \href{https://primecoin.io/primecoin-paper.pdf}{Primecoin:
  Cryptocurrency with Prime Number Proof-of-Work}.
\newblock White paper, July 2013.

\bibitem[KW26]{KW26}
I.~Komargodski and O.~Weinstein.
\newblock \href{https://doi.org/10.4230/LIPICS.ESA.2026.2}{Proofs of Useful
  Work from Arbitrary Matrix Multiplication (Invited Talk)}.
\newblock In P.~Bille, S.~Pettie, and S.~Storandt, editors, {\em 34th Annual
  European Symposium on Algorithms, {ESA} 2026, L'Aquila, Italy, August 31 -
  September 4, 2026}, volume 388 of {\em LIPIcs}, pages 2:1--2:17. Schloss
  Dagstuhl - Leibniz-Zentrum f{\"{u}}r Informatik, 2026.

\bibitem[Man18]{Man18}
N.~G. Mankiw.
\newblock {\em Principles of Microeconomics}.
\newblock Cengage Learning, Boston, MA, 8th edition, 2018.

\bibitem[Pas26]{Pas26}
R.~Pass.
\newblock \href{https://arxiv.org/abs/2606.06700}{The Economics of
  Proof-of-Useful-Work}, 2026,  arXiv:
  \href{https://arxiv.org/abs/2606.06700}{{\ttfamily 2606.06700}}.

\bibitem[PR17]{PR17}
R.~S. Pindyck and D.~L. Rubinfeld.
\newblock {\em Microeconomics}.
\newblock Pearson, 9th edition, 2017.

\bibitem[Sau92]{Sau92}
H.~D. Saunders.
\newblock \href{https://doi.org/10.5547/ISSN0195-6574-EJ-Vol13-No4-7}{The
  Khazzoom-Brookes Postulate and Neoclassical Growth}.
\newblock {\em The Energy Journal}, 13(4):131--148, 1992.

\bibitem[SKG19]{SKG19}
C.~Stoll, L.~Klaa{\ss}en, and U.~Gallersd{\"o}rfer.
\newblock \href{https://doi.org/10.1016/j.joule.2019.05.012}{The Carbon
  Footprint of Bitcoin}.
\newblock {\em Joule}, 3(7):1647--1661, July 2019.

\bibitem[Sor09]{Sor09}
S.~Sorrell.
\newblock \href{https://doi.org/10.1016/j.enpol.2008.12.003}{Jevons' Paradox
  revisited: The evidence for backfire from improved energy efficiency}.
\newblock {\em Energy Policy}, 37(4):1456--1469, 2009.

\end{thebibliography}
    \fi
\fi

\end{document}